\documentclass[12pt]{article}
\usepackage{epsfig}
\usepackage{subfigure}
\usepackage{amssymb,amsmath}
\usepackage{hyperref}
\usepackage{url}
\usepackage{setspace}

\newcommand{\bea}{\begin{eqnarray}}
\newcommand{\eea}{\end{eqnarray}}

 \makeatletter
\def\alt{\mathrel{\mathpalette\gl@align<}}
\def\agt{\mathrel{\mathpalette\gl@align>}}
\def\gl@align#1#2{\lower.6ex\vbox{\baselineskip\z@skip\lineskip\z@
\ialign{$\m@th#1\hfil##\hfil$\crcr#2\crcr\sim\crcr}}} \makeatother

\def\tst{\tilde t}
\def\ttau{\tilde \tau}

\def\tw{\widetilde W}

\newcommand{\beq}{\begin{equation}}
\newcommand{\eeq}{\end{equation}}

\begin{document}
\begin{flushright}
BA-08-19  \\
\end{flushright}
\vspace*{1.0cm}

\begin{center}
\baselineskip 20pt {\Large\bf
Soft Probes of SU(5) Unification
}
\vspace{1cm}

{\large Ilia Gogoladze$^{a,}$\footnote{ E-mail:
ilia@physics.udel.edu\\ \hspace*{0.5cm} On  leave of absence from:
Andronikashvili Institute of Physics, GAS, 380077 Tbilisi, Georgia.}
Rizwan Khalid$^{a,}$\footnote{ E-mail: rizwan@udel.edu},
Nobuchika Okada$^{b,}$\footnote{ E-mail: okadan@post.kek.jp} and
Qaisar Shafi$^{a}$ } \vspace{0.5cm}

{\baselineskip 20pt \it
$^a$Bartol Research Institute, Department of Physics and Astronomy, \\
 University of Delaware, Newark, DE 19716, USA \\
\vspace{2mm}
$^b$
Theory Group, KEK, Tsukuba 305-0801, Japan }
\vspace{.5cm}

\vspace{1.5cm} {\bf Abstract}
\end{center}

\noindent We explore the sparticle and Higgs spectroscopy of
an SU(5) inspired extension of the constrained minimal supersymmetric
standard model (CMSSM). The universal soft parameter $m_0$ is replaced
by $m_{\bar 5}$ and $m_{10}$, where $m_{\bar 5}$ and $m_{10}$ denote universal soft
scalar masses associated with fields in the five and ten dimensional
representations of SU(5). 
The special case $m_{\bar 5} \ll m_{10}$ yields a rather
characteristic sparticle spectroscopy which can be tested at the LHC.
We highlight a few benchmark points in which the
lightest neutralino saturates the WMAP bound on 
cold dark matter abundance. 

\thispagestyle{empty}

\newpage

\addtocounter{page}{-1}

\baselineskip 18pt


\section{Introduction}

With the imminent deployment of the Large Hadron Collider (LHC), a
great deal of recent theoretical research has centered around the
constrained minimal supersymmetric standard model (CMSSM). In
contrast to a generic supersymmetric standard
model which can contain more than a hundred free parameters, the
CMSSM contains just five, with three of them arising from
supersymmetry (SUSY) breaking in the hidden sector, which is then
transmitted to the visible sector through supergravity interactions~\cite{Martin:1997ns}.
These include $m_0$ and $m_{1/2}$, the
universal soft scalar and gaugino masses, and $A$, the universal coefficient of
the soft trilinear terms. The fourth parameter of the CMSSM
is $\tan \beta$, the ratio between the VEVs of the up and down type Higgs doublets in the MSSM. 
The fifth parameter is the sign of $\mu$, which we will take to
be positive in this paper.

The case for CMSSM \cite{Feldman:2008hs} becomes even more
compelling if account is taken of the apparent unification at
$M_{\rm GUT} \sim 2 \times 10^{16}$ GeV of the standard model gauge
couplings, with SUSY broken around the TeV scale. In this case, the
parameters $m_0$, $m_{1/2}$ and $A$ are specified at $M_{\rm GUT}$.
By fixing these soft SUSY breaking terms, the relic abundance of
neutralino dark matter can be predicted. The parameters determining
this relic abundance are severely constrained by the most recent
WMAP analysis~\cite{WMAP}.

Assuming that the universal soft SUSY breaking terms are prescribed
at $M_{\rm GUT}$, it is plausible that they carry some
information about the underlying GUT. For example, sparticles which
belong in a given representation of a GUT gauge group can be expected
to have identical soft masses at $M_{\rm GUT}$. The sparticle masses
at low scales are determined by the renormalization group equations (RGEs)
of the various parameters. In the CMSSM the allowed parameter space
turns out to be quite restricted after the various phenomenological
constraints are imposed.

In this paper, motivated by supersymmetric SU(5), 
we do not require identical sfermion masses at $M_{\rm GUT}$ for the $\bar 5$ and 10
matter multiplets~\cite{Profumo:2003ema}. Instead, we introduce two distinct soft mass
parameters at $M_{\rm GUT}$, denoted as $m_{\bar 5}$ and $m_{10}$.
For simplicity, we also make the plausible assumption that $m_{\bar
5}$ is also the asymptotic soft mass associated with the two Higgs
supermultiplets 5 and $\bar {5}$ of SU(5). Note that we will not
impose $b$~-~$\tau$ Yukawa unification which follows from minimal SU(5), but
which can be strongly violated, as we briefly show later, in the
presence of higher dimensional operators. This will allow us to
consider Higgs and sparticle spectroscopy without imposing
additional restrictions on the parameters.

In what follows, we will investigate a generalized CMSSM inspired by
SU(5) with the sfermion soft supersymmetry breaking (SSB) masses prescribed 
at $M_{\rm GUT}$ as follows:
\bea
 &&  m_{\tilde{D}^c} =  m_{\tilde{L}} = m_{H_u} = m_{H_d} = m_{\bar 5},
\nonumber \\
 &&  m_{\tilde{Q}} = m_{\tilde{U}^c} = m_{\tilde{E}^c} = m_{10} ,
\label{BC} \eea while the remaining parameters are the same as in the CMSSM.
Clearly, the CMSSM is realized by setting $m_{\bar 5}=m_{10}=m_0$.

Through this generalization, the resultant sparticle mass spectrum
at the weak scale differs, as one should expect, from the CMSSM one.
To see this, it is useful to examine the one-loop RGEs for the sparticle masses.
For sfermion masses in the first and second generations,
we can neglect Yukawa coupling contributions in the RGEs
and to a good approximation, the analytic expressions (see~\cite{Martin:1993zk}
and references therein) are given by
\bea
m_{\tilde{Q}}^2 & \simeq &  4.3 m_{1/2}^2 +m_{10}^2 ,   \nonumber \\
m_{\tilde{U}^c}^2 &  \simeq & 3.9 m_{1/2}^2 +m_{10}^2 ,  \nonumber \\
m_{\tilde{D}^c}^2  & \simeq & 3.9 m_{1/2}^2 +m_{\bar 5}^2 ,   \nonumber \\
m_{\tilde{L}}^2 & \simeq & 0.47 m_{1/2}^2 + m_{\bar 5}^2 ,
\nonumber \\
m_{\tilde{E}^c}^2 & \simeq &  0.15 m_{1/2}^2 + m_{10}^2 ,
\label{sfermion} \eea
where for simplicity the sfermion masses are evaluated at 1 TeV.
In these expressions, the terms proportional to $m_{1/2}$
are generated through RGEs, while $m_{\bar 5}$ and $m_{10}$ shift
the overall value of sfermion squared masses.
For  $m_{\bar 5}^2, m_{10}^2 \ll m_{1/2}^2$, the resultant sfermion
masses are largely controlled by the gaugino masses and will
be similar to the results in the CMSSM with $m_0^2 \ll m_{1/2}^2$.
On the other hand, for $m_{\bar 5}^2$ and/or $m_{10}^2 \gtrsim m_{1/2}^2$ and
 $m_{\bar 5} \neq m_{10}$, we can see a remarkable difference.
In particular, for slepton masses the gaugino mass contributions
 are not so large and as a result, the slepton mass spectrum
 can be dramatically different from the CMSSM results.

If the sparticles are discovered at the LHC and their masses measured,
we can explore the nature of SUSY breaking by extrapolating,
using RGEs, the masses towards high energies. As is easily understood
from Eq.~(\ref{sfermion}),  the first and
second generation sfermion masses from  the $\bar{\bf 5}$ and  ${\bf 10}$
will show separate unification  at $M_{\rm GUT}$.
This is in sharp contrast with the CMSSM where all sfermion masses
are unified into a single   $m_0$.
The boundary condition for the sfermion masses in the CMSSM seems more
appropriate for  a GUT model based on SO(10) or E$_6$ where all the
MSSM particles are embedded in a single representation. Thus, the
soft masses can be used as a tool to probe the structure of the underlying GUT.

Figure~1 shows a schematic picture comparing unification
 of sfermion masses in an SU(5) inspired CMSSM model with the standard CMSSM.
The  running sfermion  masses, computed to 1-loop, in the $\bar{\bf 5}$-plet
 and the ${\bf 10}$-plet are separately unified
 at $M_{\rm GUT}$ (Figure~1(a)).
The same figure for the CMSSM (or an SO(10)-like model)
 is shown in Figure~1(b), where the soft masses
 converge to a single point at $M_{\rm GUT}$. Note that in the following
analysis we employ ISAJET which computes the sparticle masses using
the full 2-loop RGEs.

\section{Phenomenological constraints and scanning procedure}

\label{ch:constraints}

We employ ISAJET~7.78 package~\cite{ISAJET} to perform random
scans over the parameter space. In this package, the weak scale
values of gauge and third generation Yukawa couplings are evolved to
$M_{\rm GUT}$ via the MSSM renormalization group equations (RGEs) in the
$\overline{DR}$ regularization scheme, where $M_{\rm GUT}$ is defined to
be the scale at which $g_1=g_2$. We do not enforce an exact
unification of the strong coupling $g_3=g_1=g_2$ at $M_{\rm GUT}$, since
a few percent deviation from unification can be assigned to unknown
GUT-scale threshold corrections~\cite{Hisano:1992jj}. At $M_{\rm GUT}$,
the boundary conditions presented in Eq.~(\ref{BC}) are imposed and all the SSB
parameters, along with the gauge and Yukawa couplings, are evolved
back to the weak scale $M_{\rm Z}$. In the evaluation of  Yukawa couplings
the SUSY threshold corrections~\cite{Pierce:1996zz} are taken into
account at the common scale $M_{\rm SUSY}= \sqrt{m_{\tst_L}m_{\tst_R}}$.
The entire parameter set is iteratively run between $M_{\rm Z}$ and
$M_{\rm GUT}$ using the full 2-loop RGEs until a stable solution is
obtained. To better account for leading-log corrections, one-loop
step-beta functions are adopted for gauge and Yukawa couplings, and
the SSB parameters $m_i$ are extracted from RGEs at multiple scales
$m_i=m_i(m_i)$.  The RGE-improved 1-loop effective potential is
minimized at an optimized scale $M_{\rm SUSY}$, which effectively
accounts for the leading 2-loop corrections. Full 1-loop radiative
corrections are incorporated for all sparticle masses.

The requirement of radiative electroweak symmetry breaking
(REWSB)~\cite{Ibanez:1982fr} puts an important theoretical
constraint on the parameter space. Another important constraint
comes from limits on the cosmological abundance of stable charged
particles~\cite{Yao:2006px}. This excludes regions in the parameter space
where charged SUSY particles, such as $\ttau_1$ or $\tst_1$, become
the lightest supersymmetric particle (LSP). We accept only those solutions
for which the neutralino is the LSP.

We have performed random scans for the following parameter range:

\begin{eqnarray}
&& 0\leq m_{\bar 5} \leq 5\, \textrm{TeV}, \nonumber \\
&& 0\leq m_{10} \leq 5\, \textrm{TeV},  \nonumber \\
&& 0\leq m_{1/2} \leq 2\, \textrm{TeV}, \nonumber  \\
&& A_0= 0.5\, \textrm{TeV},\ 0,\ -1\, \textrm{TeV},\  -2\, \textrm{TeV}, \nonumber \\
&& \tan \beta =5,\ 10,\ 30,\ 50\ \textrm{and}\ 55,
\label{ppp1}
\end{eqnarray}
with $\mu >0$ and $m_t = 172.6$~GeV \cite{Group:2008nq}.

After collecting the data, we use the IsaTools package~\cite{Baer:2002fv}
to implement the following phenomenological constraints:
\begin{eqnarray}
&& m_{\tw_1}~{\rm (chargino~mass)} \geq 103.5~{\rm GeV} \qquad {\rm \cite{Yao:2006px}}, \nonumber \\
&& m_h~{\rm (lightest~Higgs~mass)} \geq 114.4~{\rm GeV} \qquad {\rm \cite{Schael:2006cr}}, \nonumber \\
&& BF(B_s \rightarrow \mu^+ \mu^-)< 5.8 \times 10^{-8} \qquad {\rm \cite{:2007kv}}, \nonumber \\
&& \Omega_{\rm CDM}h^2 = 0.111^{+0.011}_{-0.015} \qquad (2\sigma) \qquad {\rm \cite{WMAP}}, \nonumber \\
&& 2.85 \times 10^{-4} \leq Br(b \rightarrow s \gamma)\leq 4.24 \times 10^{-4} \qquad (2\sigma) \qquad {\rm \cite{Barberio:2007cr}}, \nonumber \\
&& 3.4 \times 10^{-10}\leq \Delta a_{\mu} \leq 55.6 \times 10^{-10}~
\qquad (3\sigma) \qquad {\rm \cite{Bennett:2006fi}}.
\end{eqnarray}
We have applied the constraints
from experimental data successively on the data that we acquired from ISAJET. First
we apply the constraint on the $BR(B\rightarrow \mu^+ \mu^-)$, then the constraint
on the chargino mass, followed by the WMAP upper bound on the relic density of cold dark
matter. The constraint from $BR(B\rightarrow X_{s} \gamma)$ is then taken into consideration,
followed by the constraint on $\Delta a_\mu$. Finally, we apply also the
lower bound on the dark matter relic abundance. The data is then plotted showing the
successive application of each of these constraints.

The color coding is explained below, as well as in
Figure~\ref{fund-t30a0} caption.

\begin{itemize}

\item Black: Points excluded by the LEP 2 bound on the Higgs mass.
\item Gray: Points that satisfy  $BR(B\rightarrow \mu^+ \mu^-)$ and the chargino mass bound.
\item Light Green: Points that satisfy the WMAP upper bound on dark matter relic abundance.
\item Dark Green: Points that satisfy both upper and lower bounds on dark matter relic abundance.
\item Light and Dark Blue: Points that satisfy $BR(B\rightarrow X_{s} \gamma)$. Light blue points only
satisfy the lower bound on dark matter relic abundance while dark blue ones satisfy both upper and lower bounds.
\item Orange and Red: Points that satisfy the constraint from $\Delta a_\mu$. Orange points satisfy only the
lower bound on dark matter relic density while dark blue ones satisfy both upper and lower bounds.

\end{itemize}

Thus, behind every red point, there is a dark blue, a dark green, and a gray point. Likewise,
behind every orange point there is a light blue, a light green and a gray point.

\section{$b$~-~$\tau$ (Non)~-~Unification}

Before proceeding further, let us briefly discuss the issue of
$b$~-~$\tau$ Yukawa unification. In minimal SU(5) with a single
$5+\bar 5$ pair of Higgs, the Yukawa couplings of $b$ and $\tau$ are
equal at $M_{\rm GUT}$. However, due to potentially large radiative
corrections to the $b$ mass~\cite{Hall:1993gn}, this asymptotic relation does not
lead to a satisfactory prediction for the $b$ quark mass without
making additional assumptions about the soft masses, $A$ terms and
$\tan \beta$. Thus, in this paper we will not
impose $b$~-~$\tau$ Yukawa unification at $M_{\rm GUT}$. Indeed,
one could expect this naive unification relation to be modified for a
number of reasons. Consider the following $b$ and $\tau$ Yukawa
couplings (at $M_{\rm GUT}$)
\begin{eqnarray}
\label{adjointeqn}
y~10_{\alpha \beta} \, \bar 5^{\alpha}\, \bar 5^{\beta}
+\frac{{\lambda}_1}{\Lambda}\, 10_{\alpha \beta}\,
\Sigma_{\gamma}^{\beta}\, \bar 5^{\gamma}_f \, \bar 5^{\alpha}_H
+\frac{{\lambda}_2}{\Lambda}10_{\alpha \beta}
\Sigma^{\beta}_{\gamma}\, \bar 5^{\alpha}_f \, \bar 5^{\gamma}_H,
\end{eqnarray}
where Greek letters denote SU(5) indices, and
the SU(5) adjoint Higgs $\Sigma$ develops a VEV ($v$) which breaks SU(5) to MSSM.
We have included dimension five terms in Eq.~(\ref{adjointeqn})
to show how departure from $b$~-~$\tau$ unification can arise. Indeed, 
such higher order terms have previously been used~\cite{Ellis:1979fg,Panagiotakopoulos:1984wf} 
to modify the `bad' Yukawa relations
for the first two families predicted in minimal SU(5). The cutoff scale ${\Lambda}$
can be the reduced Planck mass $M_{PL} = 2.4 \times 10^{18}~{\rm GeV}$,
or it can be a superheavy mass scale of order
$M_{\rm GUT}$, associated with suitable vector-like particles.
[Integrating out these latter states should yield the desired dimension five
operators].

From Eq.~(\ref{adjointeqn}), we find
\begin{eqnarray}
&& y_b   = y + {\lambda}_1 (2v)/{\Lambda} + {\lambda}_2 (-3v)/{\Lambda}  \nonumber \\
&& y_{\tau} = y + {\lambda}_1 (-3v)/{\Lambda} + {\lambda}_2(-3v)/{\Lambda}. \label{ybytaueqn}
\end{eqnarray}
\noindent
For simplicity let us set $\lambda_2$ = 0, so that
\beq y_b   = y + 2 y'  ~ {\rm and}~   y_\tau = y - 3 y', 
\label{ybtaueqnmodified}\eeq
where $y' = \lambda_1 v/{\Lambda}$.  These equations allow us to express $y$
and $y'$ in terms of $y_b$ and $y_\tau$:
\begin{equation}
y  = (3 y_b + 2 y_\tau)/5 ~{\rm and}~  y' = (y_b -y_\tau)/5.
\label{yyprimeeqn}
\end{equation}

\noindent
With $y_b$ and $y_\tau$ determined in conjunction with the various phenomenological
constraints, the expressions in Eq. (\ref{yyprimeeqn}) provide the
appropriate values for $y$ and $y'$. Much of the viable parameter space
obtained in this paper does not respect $b$~-~$\tau$ unification (see Figure~\ref{yb_ybtau_tanb}).

\section{Results}

Figures~\ref{fund-t30a0} shows the results in the ($m_{\bar 5},
m_{10}$), ($m_{1/2},m_{10}$) and ($m_{1/2},m_{\bar 5}$) planes for
$\tan \beta = 30, \, \,A_0=0$   and  $\mu > 0$. In the ($m_{\bar 5},
m_{10}$) plane, lines corresponding to fixed ratios of
$m_{10}/m_{\bar 5}=0.25,  1$ and $5.1$ are plotted in
Figure~\ref{fund-t30a0} as a reference. Interestingly, we can see
that the allowed region satisfying all the constraints is very
limited, with wide excluded regions appearing in white. In the white
region between the lines $1 \lesssim m_{10}/m_{\bar 5} \lesssim
5.1$, the neutralino LSP is bino-like and its relic abundance is too
large to be consistent with WMAP data. The lighter stau is the LSP
(which can even be tachyonic) in the white region below the line
$m_{10}/m_{\bar 5} \simeq 0.25$. The white region for $m_{\bar 5}
\gtrsim 2.5$ TeV and below the line $ m_{10}/m_{\bar 5} = 1 $ is
excluded since no radiative electroweak symmetry breaking occurs
there. In the ($m_{1/2},m_{10}$) and ($m_{1/2},m_{\bar 5}$) planes,
the region with $m_{1/2} \lesssim 0.15$ TeV is excluded since no
radiative electroweak symmetry breaking occurs there.  The region
with a small $m_{10}$ in the ($m_{1/2},m_{10}$) plane is excluded
because the lighter stau is the LSP (which can even be tachyonic).
Similar remarks hold for Figures~\ref{fund-t30a1}, \ref{fund-t50a0},
\ref{fund-t50a1}, \ref{fund-t55a0}, and  \ref{fund-t55a1} which show
analogous plots for different values of $\tan \beta$ and $A_0$.

Figure~\ref{c81550m5mhf} shows the results in the ($m_{1/2},m_{\bar
5}$) plane for several fixed values of $m_{10}/m_{\bar 5}$, with
$\tan \beta=50$, $A_0=0$ and $\mu > 0$. Let us examine
Figure~\ref{c81550m5mhf}(d) in  more detail, which exhibits the case
$m_{10} \gg m_{\bar 5}$. The red and blue dots show the allowed
parameter sets (red dots are favored by the muon $g-2$ experiments),
and there are two branches for the allowed regions. The white
regions in upper-left and in lower-right corners are excluded since
the lighter stau is the LSP (it can even be tachyonic). The central
region is excluded due to  over-abundance of the neutralino LSP. In
both the allowed branches, the neutralino LSP is bino-like and
quasi-degenerate with the lighter stau. Therefore, the allowed
region is the so-called co-annihilation region in the CMSSM setup.
However, there is a crucial difference in the composition  of the
lighter stau. In the lower branch, the lighter stau is mostly  the
right-handed stau as in the CMSSM, while it is mostly a left-handed
stau in the upper branch. This is because in the upper branch
$m_{10}$ is large,  so that the right-handed stau in the SU(5) ${\bf
10}$-plet is heavy. As a result, the relic abundance of the
bino-like neutralino can be consistent with the WMAP data through
the co-annihilation process with the mostly left-handed stau. This
scenario is absent in the CMSSM, but is very characteristic for our
SU(5) inspired MSSM.

Next we select benchmark points from each of the allowed branches
 and compare the sparticle and Higgs boson masses with those in the CMSSM.
The results are shown in Table~1, together with
 the neutralino relic abundance and the spin-independent cross
 sections for   neutralino - nucleon (proton and neutron) scattering.
For comparison with the CMSSM, we fix $m_0=m_{10}$, with all other
parameters the same.  We can see sizable differences in the
sparticle mass spectra between the SU(5) model and the CMSSM.
In particular, the difference is remarkable for the parameter
 set in the upper branch because of $m_{10}=m_0 > m_{1/2} \gg m_5$.
The sfermions in the $\bar{\bf 5}$ representation of  SU(5)
 are much lighter than the corresponding sparticles in the CMSSM.
The neutralino-nucleon cross sections are a few orders of
 magnitude smaller than the exclusion limits given by
 the current experiments for direct dark matter  detection
 such as  CDMS \cite{CDMS} and XENON10 \cite{XENON}.

Let us display other characteristic results of our model with a
different set of input parameters, in particular a non-zero $A_0$.
Two cases are shown in Table~2, together with a corresponding CMSSM
 result for $m_0=m_{10}$.
There is no phenomenologically viable solution  in the CMSSM
 corresponding to the second case (the 4th column), since a
 lighter stop is found to be tachyonic in the CMSSM.
In both SU(5) examples, the input $m_{1/2}$ values are
relatively  small, and as a result, all gauginos (gluino, gaugino-like chargino,
 and gaugino-like neutralinos) are relatively  light compared to squarks.
In the CMSSM, a small $m_{1/2}$ input, say, $m_{1/2} \lesssim 300$ GeV
 is excluded by the LEP 2 bound on the lightest Higgs boson mass,
 unless  $m_0$  is large, $m_0 \gtrsim 1$ TeV.
This is because the radiative corrections to the lightest Higgs boson
 mass via a heavy stop are necessary to make the Higgs boson mass
 (which is lighter than Z-boson mass at tree level) higher
 than the LEP 2 bound.
Although the situation is the same for our model and the mass of
 ${\bf 10}$-plet including stops should be large,
 sfermions in $\bar{\bf 5}$-plet can be (much) lighter than
 those in ${\bf 10}$-plet, and still be  consistent with the
phenomenological constraints.

In both cases, the relic abundance of the neutralino LSP
 matches the WMAP data and thus the neutralino can be the
 dominant component of cold dark matter in the present universe.
Since the neutralino is bino-like, the co-annihilation process
 with a quasi-degenerate tau-sneutrino plays a crucial role
in order to yield an appropriate relic abundance. Again, this case
is not realized   in the CMSSM. In general, if the  input values of
$m_{1/2}$ and $m_{\bar 5}$ are relatively
 small compared to  $m_{10}$, the tau-sneutrino is likely to be the NLSP.

Table~2 shows that some of the colored sparticles can be light:
Gluino and right-handed down-type squarks are light because
of the small $m_{1/2}$ and $m_{\bar 5}$ input values.
The large value of $A_0$ input leads to a large mass splitting
 in stop mass eigenvalues.
In the second case (the 4th column), in particular,
 the lightest stop is remarkably light.
If gluinos are copiously produced at the LHC,
 gluino decays into the third generation squarks provide
 top quarks in the final state.
Studies of this process may reveal a sparticle nature
 related  to the third generation squarks \cite{stop}.
For the second SU(5) case (the 4th column in Table~2),
 the lighter stop is sufficiently light, so that its dominant decay mode
 is into a top quark and neutralino LSP, while the branching
 ratio to this process in the CMSSM is small.
This process is certainly worth investigating at the LHC.

Let us briefly mention the A~-~funnel region that we observe in our
model. Figure~\ref{funnel} shows plots in the
($m_{\tilde{\chi}_1^0},m_A$) plane with $A_0 = 0$ and $\tan \beta =
10, 30, 50~ {\rm and}~ 55$. Also shown in each case is the line $m_A
= 2~m_{\tilde{\chi}_1^0}$ from which we find that the A~-~funnel
region appears for $\tan \beta = 50 ~ {\rm and} ~55$, where neutralinos can
annihilate via the A and H Higgs bosons.

Finally, as discussed earlier, we have not imposed $b$~-~$\tau$ Yukawa 
unification in this paper. For completeness, we plot in Figure~\ref{yb_ybtau_tanb}
the ratio $y_b/ y_{\tau}$, evaluated at $M_{\rm GUT}$, versus $\tan \beta$. For the 
parameter space we have considered this ratio turns out to be $\lesssim 3/4$.
It is amusing to note that the value  2/3 for this ratio, which contains 
phenomenologically viable points, can be obtained from Eq.~(\ref{ybtaueqnmodified}) by 
setting $y \approx 0$.

\section{Conclusion}

We have generalized the CMSSM parameterization
 to one possibly more suited for SU(5) models by replacing
 the universal sfermion mass $m_0$ with  two independent
 sfermion masses, $m_{\bar 5}$ and $m_{10}$, corresponding to
 the five- and ten-dimensional
 representations of SU(5).
By imposing  a variety of phenomenological constraints,
 we have identified the allowed parameter space.
For points chosen from the allowed parameter space,
 we have shown that the resultant sparticle mass spectrum
 can be quite different from the one obtained in the CMSSM,
and this difference can be tested at  the LHC. With the  sparticle masses
precisely measured, we can employ them  as a tool
 to probe the underlying  GUT using  RGEs.
This SU(5) inspired version of the CMSSM shows separate  unification 
of sfermion masses in the  {$\bar 5$}- and 10-dimensional representations.

\section*{Acknowledgments}
We thank Howie Baer and Shahida Dar for helpful discussions.
N.O. would like to thank the Particle Theory Group
of the University of Delaware for hospitality during his visit.
He would also like to thank the Maryland Center
for Fundamental Physics, and especially Rabindra N. Mohapatra
for their hospitality and financial support during his stay.
This work is supported in part by
the DOE Grant \# DE-FG02-91ER40626 (I.G. and Q.S.),
GNSF grant 07\_462\_4-270 (I.G.),
the National Science Foundation Grant No. PHY-0652363 (N.O.),
and the Grant-in-Aid for Scientific Research from the Ministry
of Education, Science and Culture of Japan, \#18740170 (N.O.).


\newpage
\begin{figure}[htbp,width=20cm, height=6cm]
\begin{center}
\subfigure[]
{\includegraphics[width=10cm]{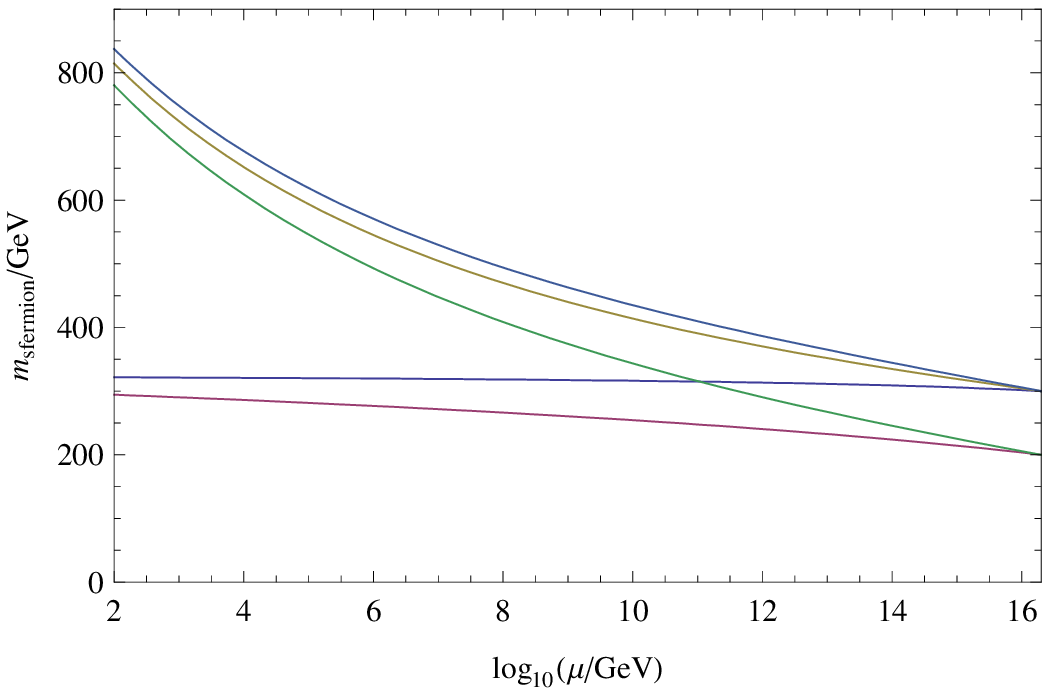}\label{Fig1a}}
\subfigure[]
{\includegraphics[width=10cm]{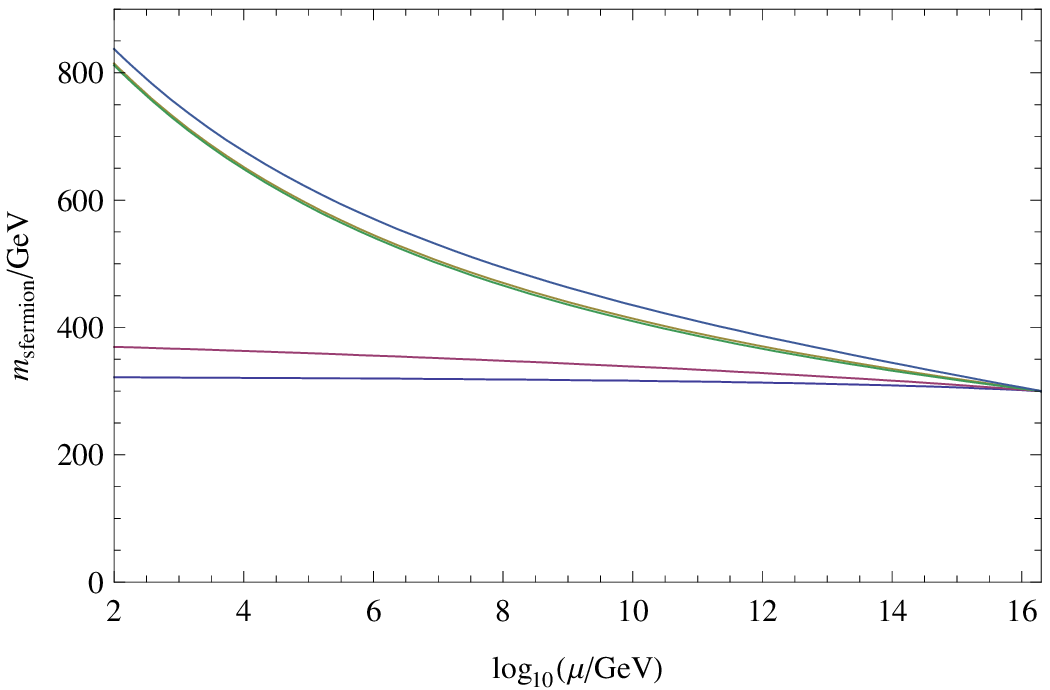}\label{Fig1b}}
\end{center}
\caption{ Evolution of the first two family sfermion
masses ($m_{\tilde{Q}}$, $m_{\tilde{U}^c}$, $m_{\tilde{D}^c}$,
$m_{\tilde{E}^c}$ and $m_{\tilde{L}}$, from top to bottom) in
(a) SU(5), (b) CMSSM (where  $m_{\tilde{U}^c} \simeq m_{\tilde{D}^c}$). 
Here $m_{1/2}=300$ GeV, $\tan \beta =30$, $A_0=0$ and $\mu>0$ for both cases, and $m_{\bar 5}=200$ GeV,
 $m_{10}=300$ GeV for SU(5), while $m_{0}=300$ GeV for CMSSM.
}
\end{figure}

\begin{figure}

\centering
\includegraphics{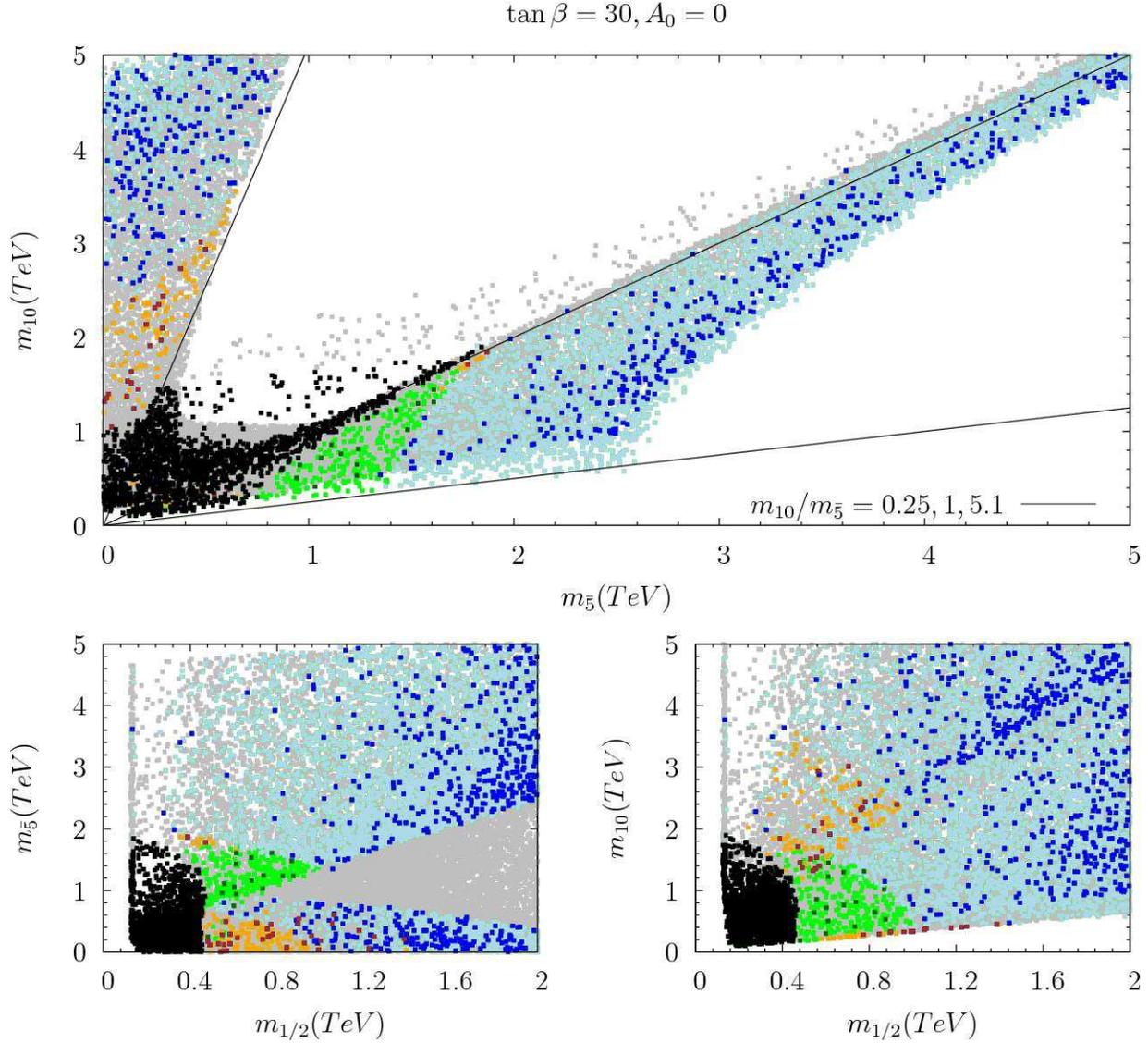}

\caption{Plots in ($m_{10}$,$m_{\bar 5}$), ($m_{\bar
5}$,$m_{1/2}$), and ($m_{10}$,$m_{1/2}$) planes for $\tan\beta = 30,
A_0 = 0, {\mu}>0$. The black region is excluded by the LEP 2 bound
on the Higgs mass. Gray points satisfy constraints from
$BR(B\rightarrow \mu^+ \mu^-)$ and the chargino mass bound.  Light
green points satisfy the WMAP upper bound on dark matter relic
abundance. Dark green points satisfy both the upper and lower
bounds on dark matter relic abundance. Light and dark blue points
satisfy $BR(B\rightarrow X_{s} \gamma)$. Light blue points only satisfy
the lower bound on dark matter relic abundance, while dark blue ones
satisfy both upper and lower bounds.  Orange and red points satisfy
the constraint from $\Delta a_\mu$. Orange points satisfy only the lower
bound on dark matter relic density, while dark blue ones satisfy both the
upper and lower bounds.\label{fund-t30a0}}

\end{figure}

\begin{figure}
\centering
\includegraphics{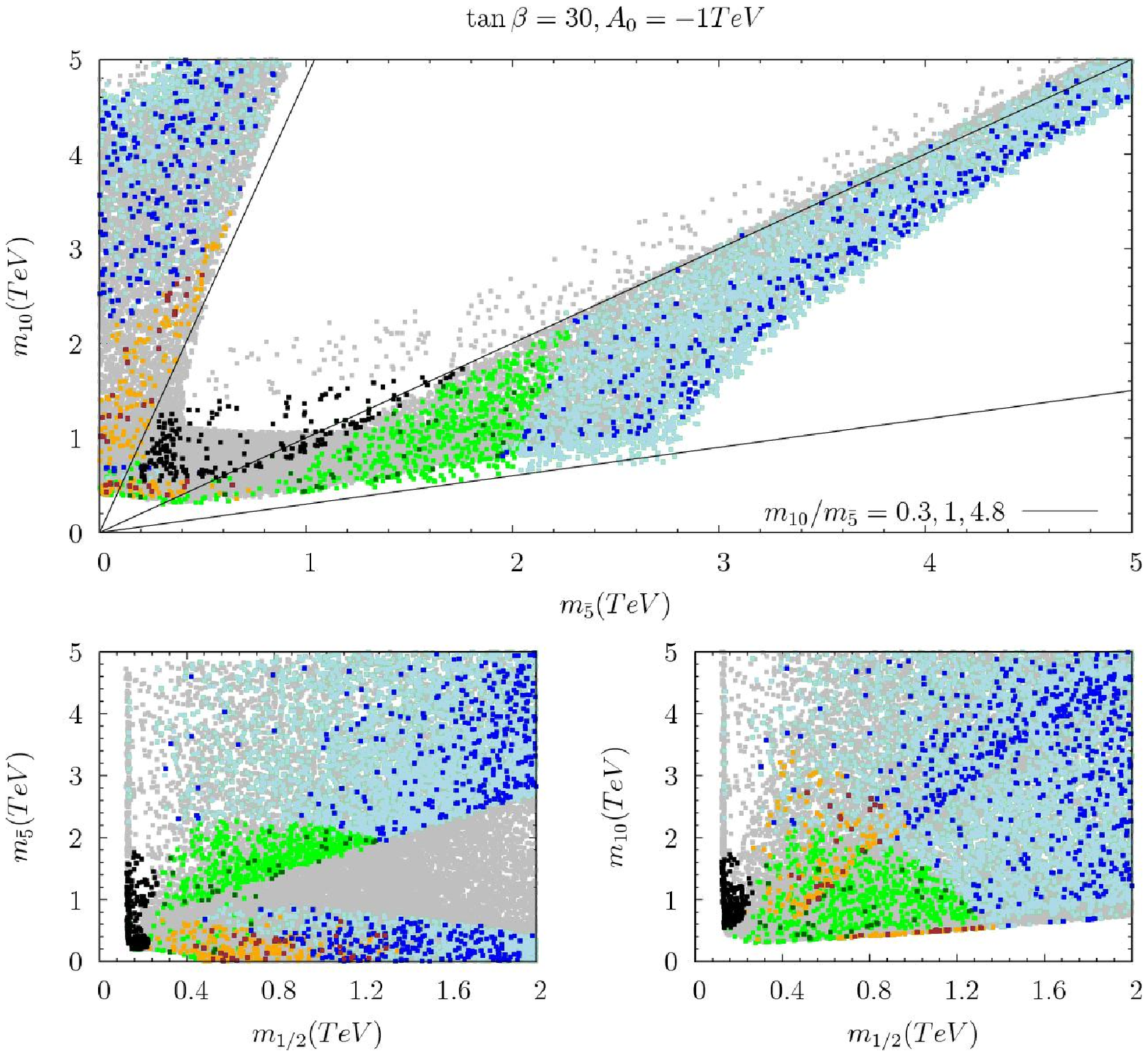}

\caption{Plots in ($m_{10}$,$m_{\bar 5}$), ($m_{\bar 5}$,$m_{1/2}$),
($m_{10}$,$m_{1/2}$) planes for $\tan\beta = 30, A_0 = -1 {\rm TeV}, {\mu}>0$. Color
coding same as in Figure \ref{fund-t30a0}.\label{fund-t30a1}}

\end{figure}

\begin{figure}
\centering
\includegraphics{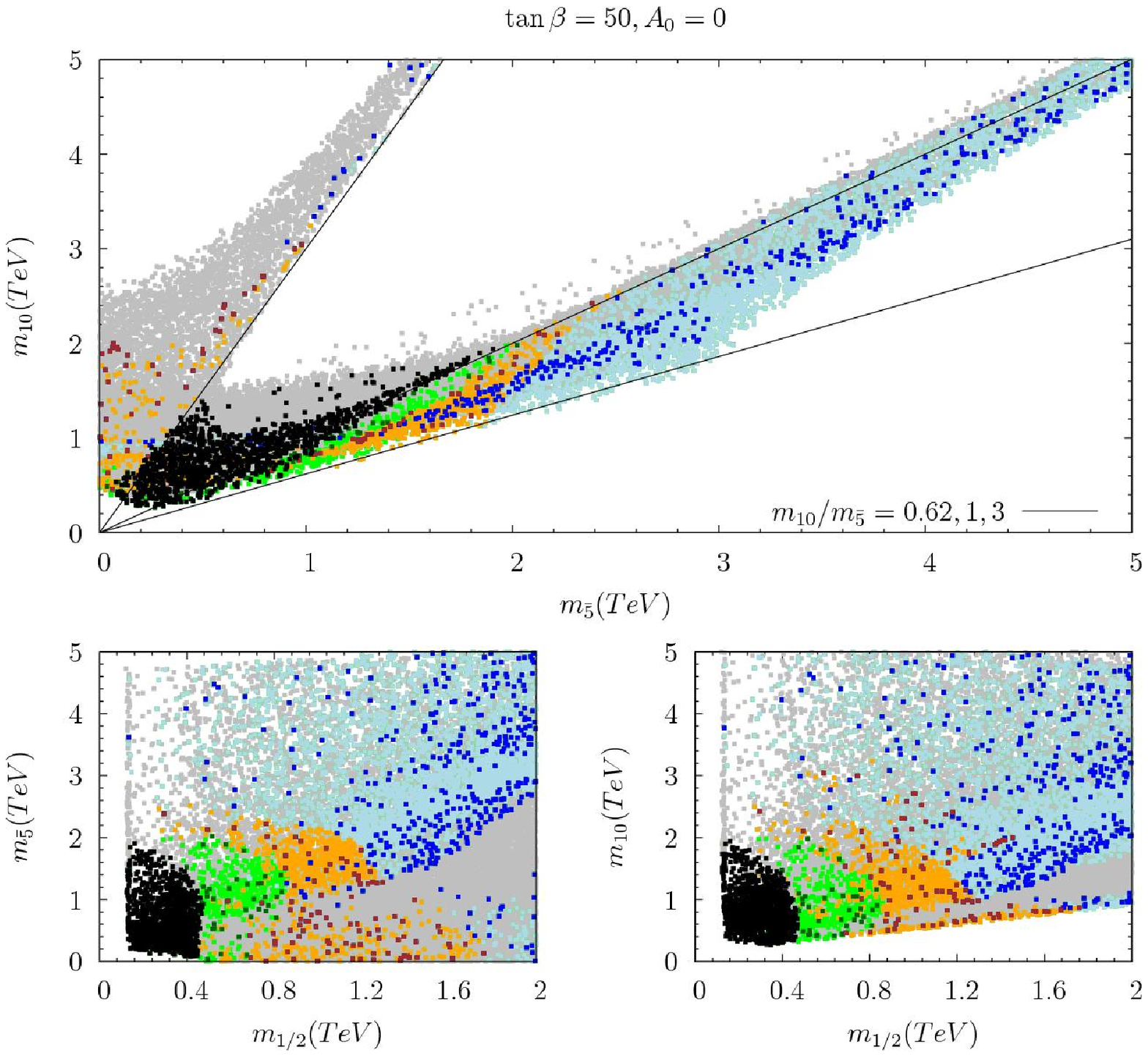}

\caption{Plots in ($m_{10}$,$m_{\bar 5}$), ($m_{\bar 5}$,$m_{1/2}$),
($m_{10}$,$m_{1/2}$) planes for $\tan\beta = 50, A_0 = 0, {\mu}>0$. Color
coding same as in Figure \ref{fund-t30a0}.\label{fund-t50a0}}

\end{figure}

\begin{figure}
\centering
\includegraphics{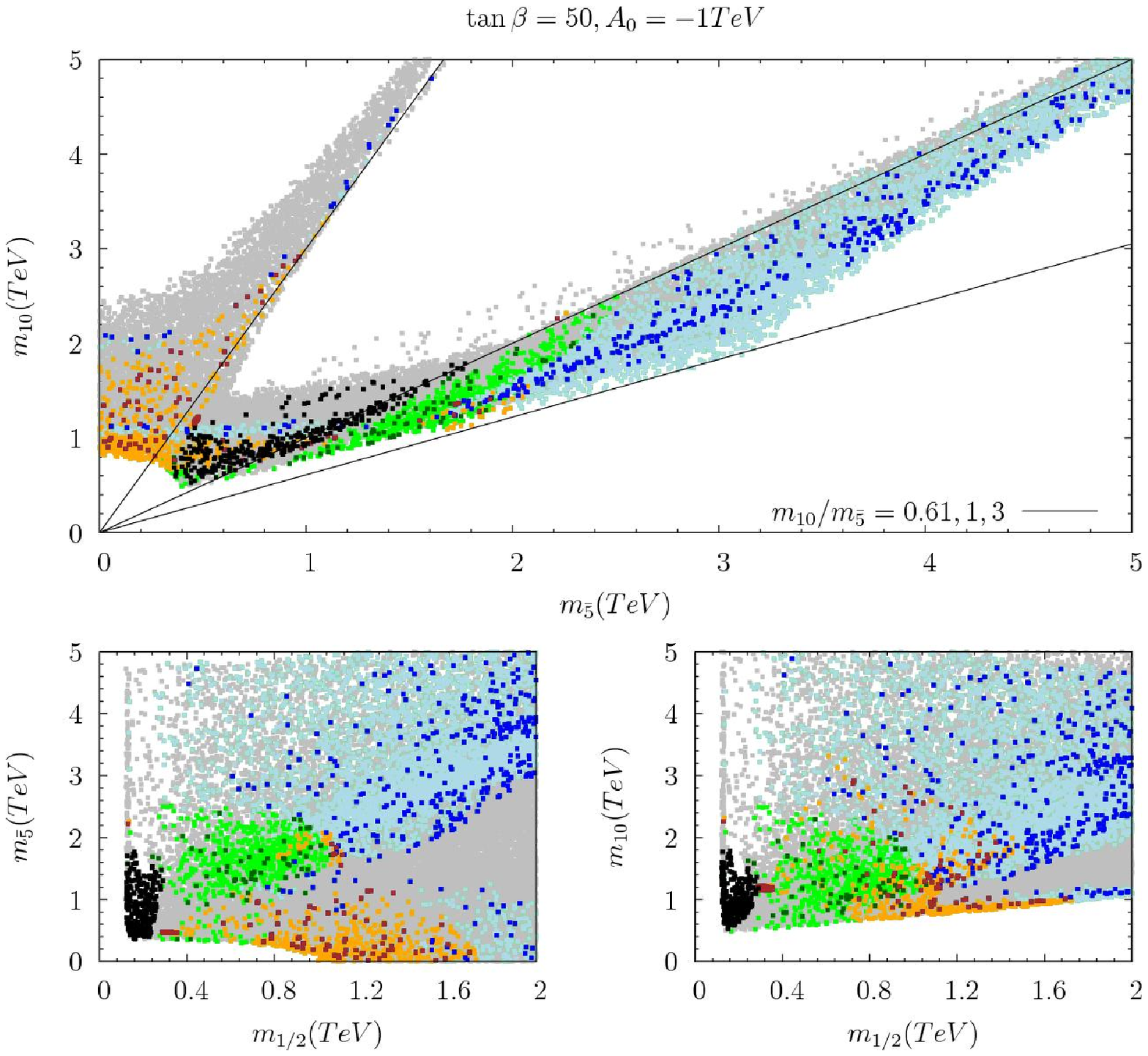}

\caption{Plots in ($m_{10}$,$m_{\bar 5}$), ($m_{\bar 5}$,$m_{1/2}$),
($m_{10}$,$m_{1/2}$) planes for $\tan\beta = 50, A_0 = -1 {\rm TeV}, {\mu}>0$. Color
coding same as in Figure \ref{fund-t30a0}.\label{fund-t50a1}}

\end{figure}

 \begin{figure}
 \centering
 \includegraphics{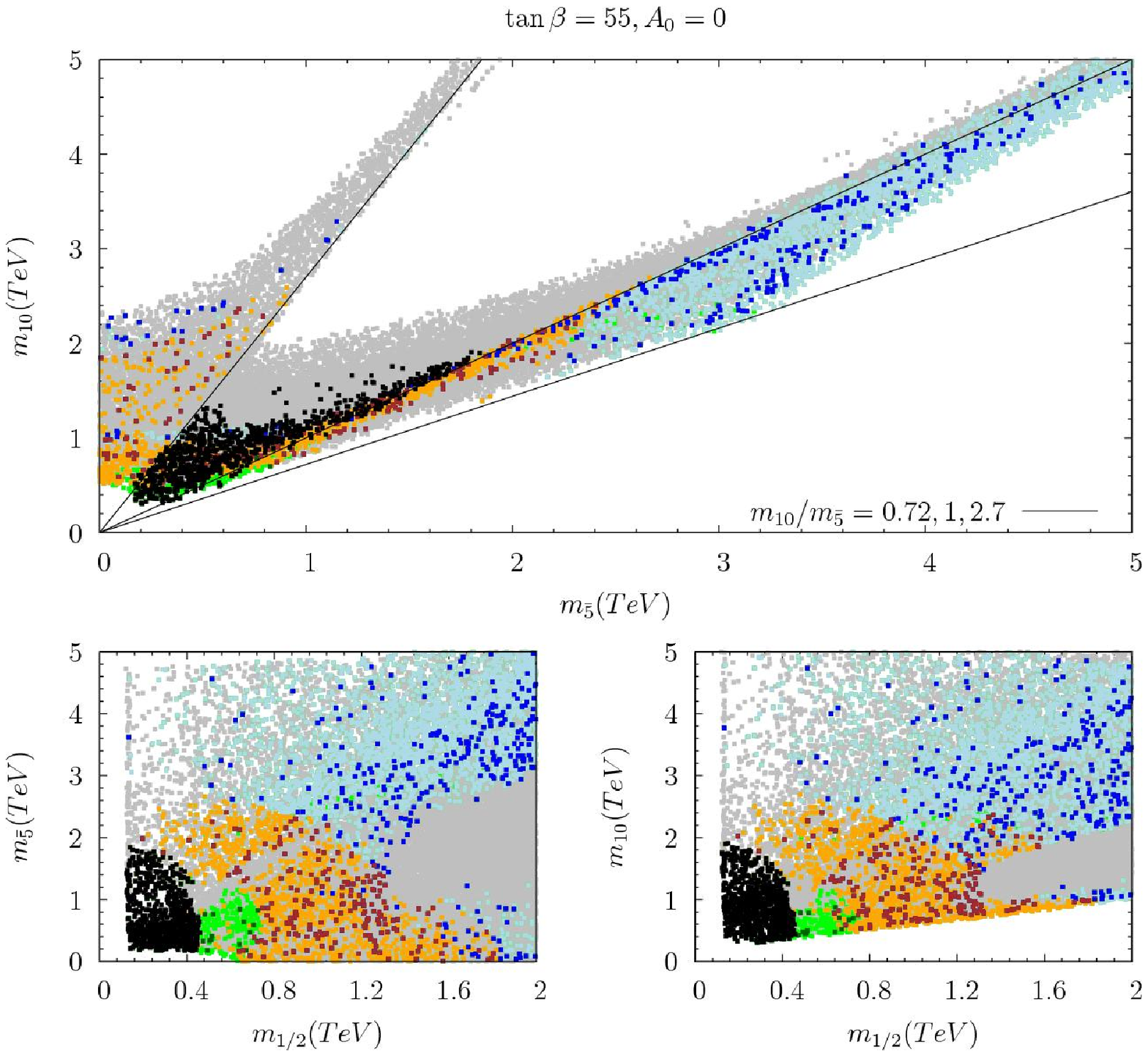}

 \caption{Plots in ($m_{10}$,$m_{\bar 5}$), ($m_{\bar 5}$,$m_{1/2}$),
 ($m_{10}$,$m_{1/2}$) planes for $\tan\beta = 55, A_0 = 0, {\mu}>0$. Color
 coding same as in Figure \ref{fund-t30a0}.\label{fund-t55a0}}

 \end{figure}

 \begin{figure}
 \centering
 \includegraphics{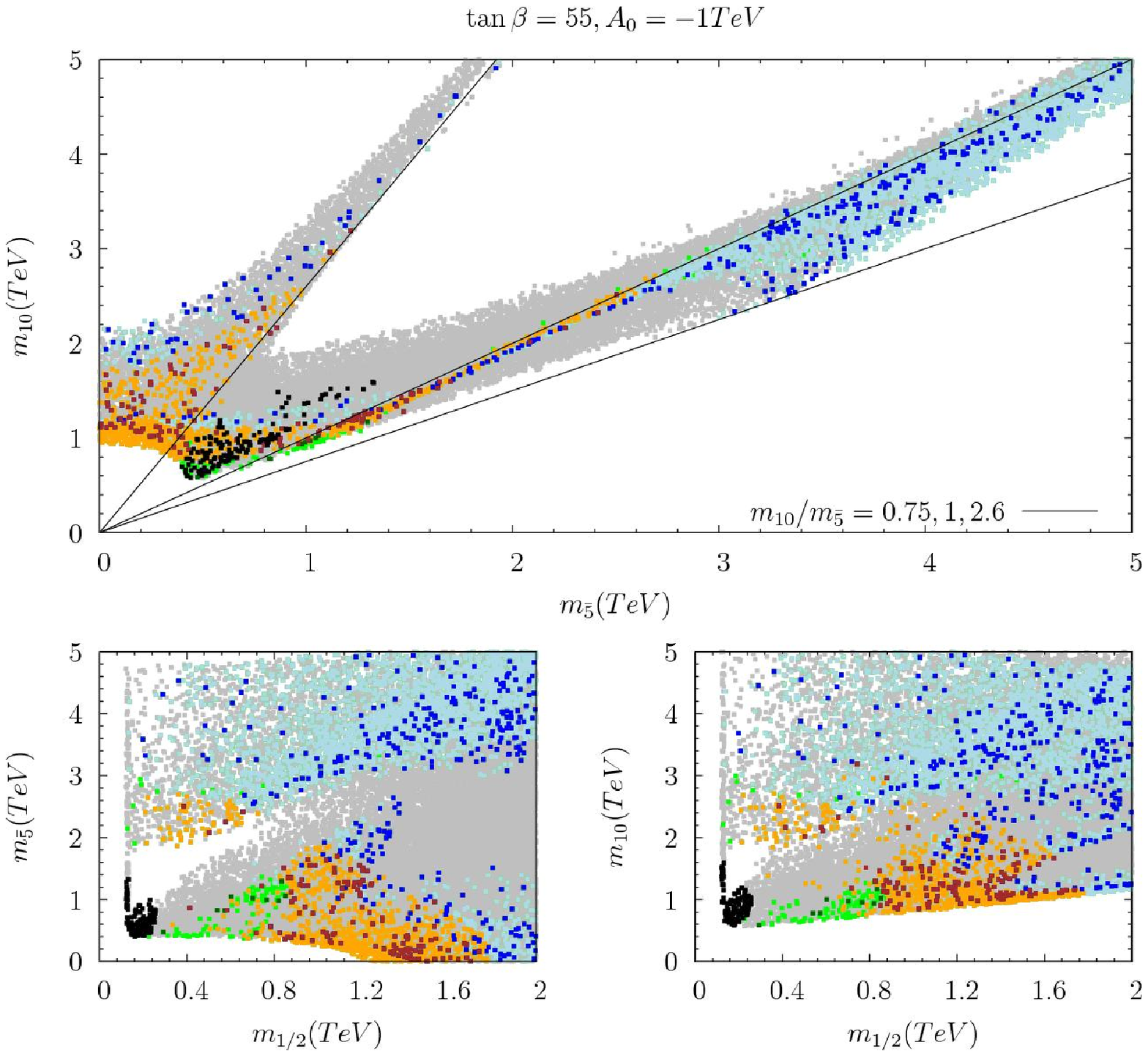}

 \caption{Plots in ($m_{10}$,$m_{\bar 5}$), ($m_{\bar 5}$,$m_{1/2}$),
 ($m_{10}$,$m_{1/2}$) planes for $\tan\beta = 55, A_0 = -1 {\rm TeV}, {\mu}>0$. Color
 coding same as in Figure \ref{fund-t30a0}.\label{fund-t55a1}}

 \end{figure}

\begin{figure}
\centering
\includegraphics{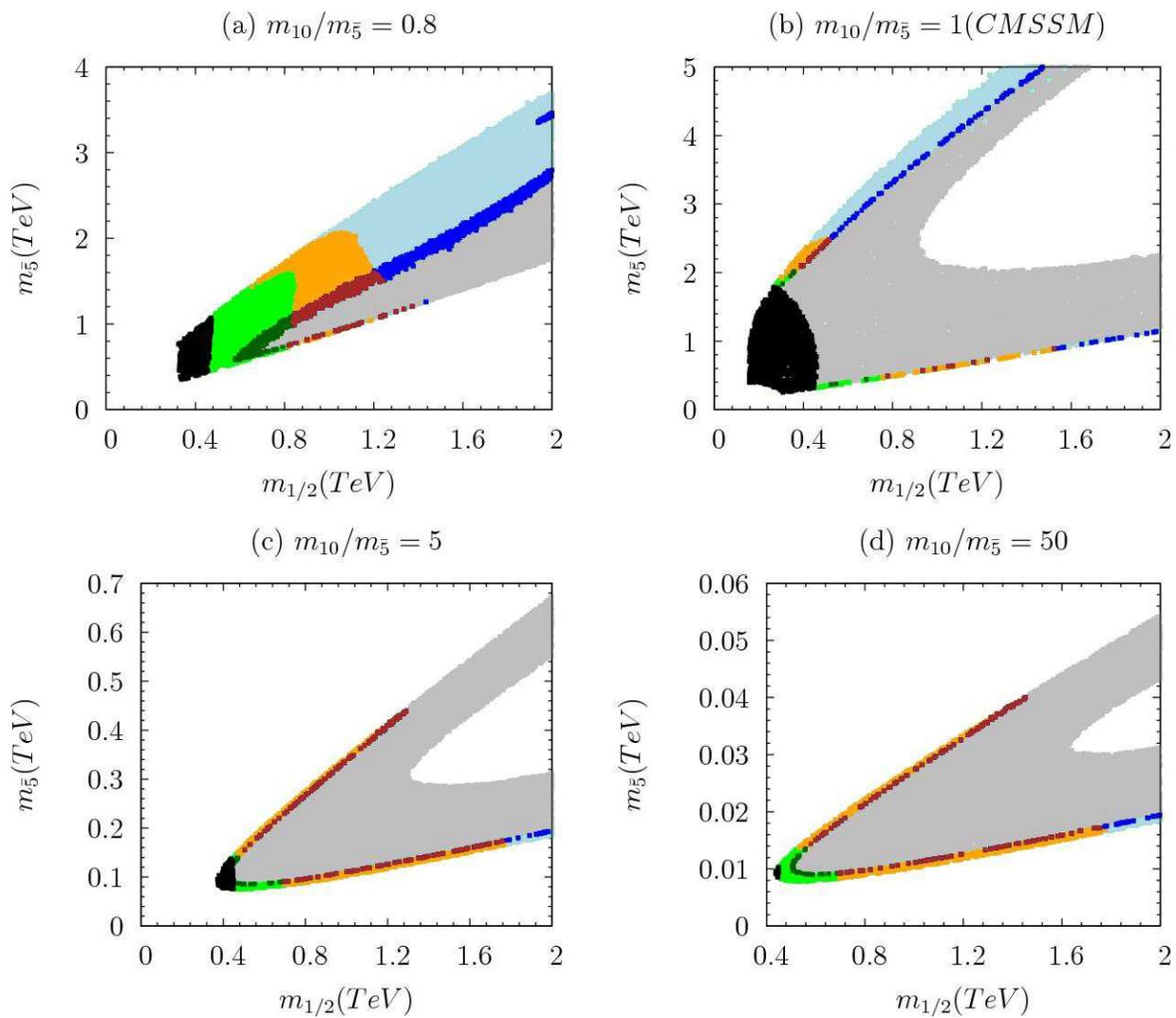}

\caption{($m_{\bar 5}$,$m_{1/2}$) plane for $\tan\beta = 50, A_0=0, {\mu}>0$ with
$m_{10}/m_{\bar 5}$ = 0.8, 1 (CMSSM), 5, and 50. Color
coding same as in Figure \ref{fund-t30a0} \label{c81550m5mhf}}

\end{figure}

\begin{figure}
\centering
\includegraphics{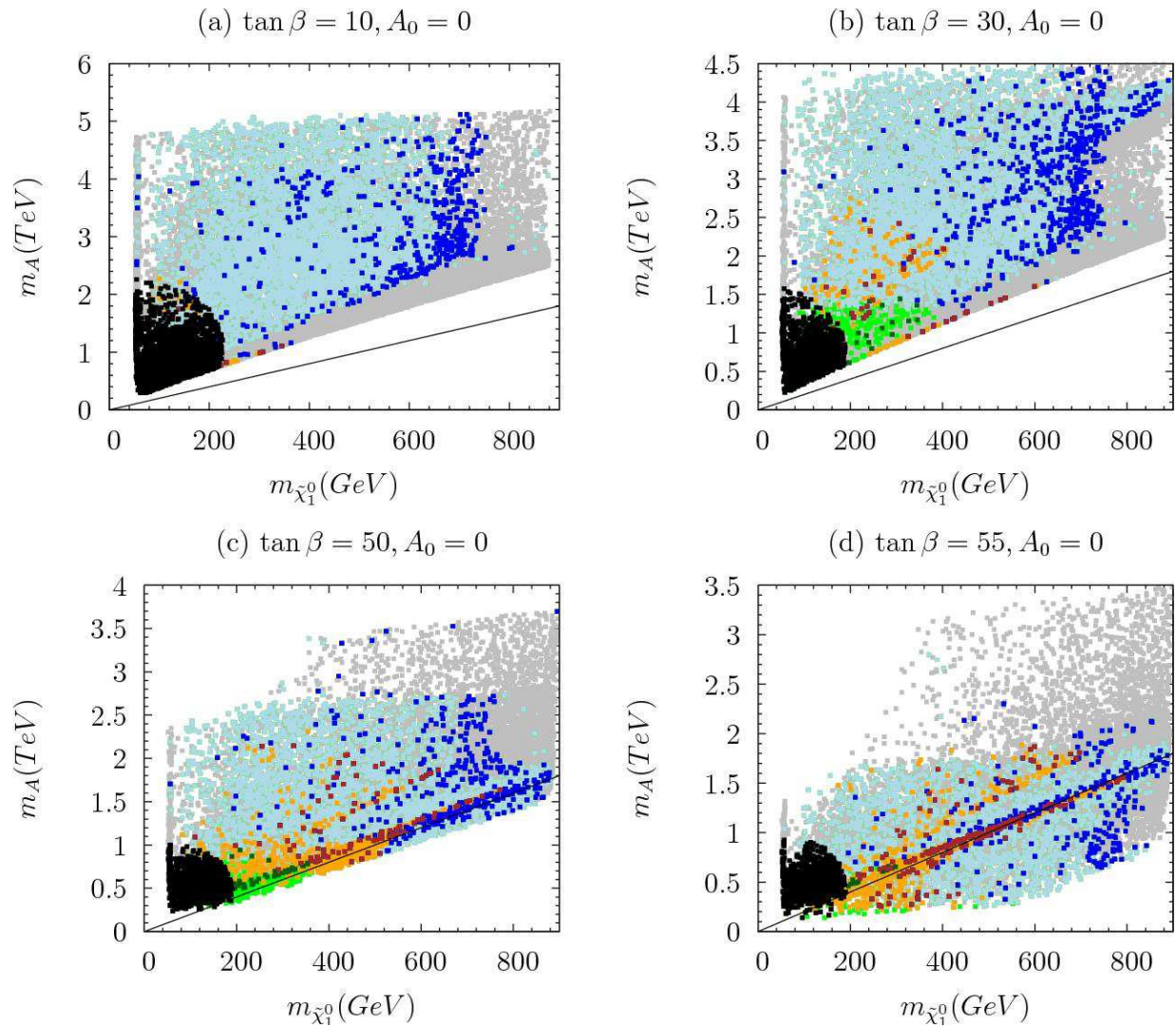}

\caption{Plots in ($m_A$,$m_{\tilde{\chi}_1^0}$) plane with $A_0
= 0,  {\mu}>0$ and $\tan \beta = 10, 30, 50~ {\rm and}~ 55$. Also
shown in each case is the line $m_A = 2~m_{\tilde{\chi}_1^0}$
which indicates existence of the A-funnel region in (c) and (d). Color 
coding same as in Figure~\ref{fund-t30a0}.\label{funnel}}

\end{figure}

\begin{figure}
\centering
\includegraphics{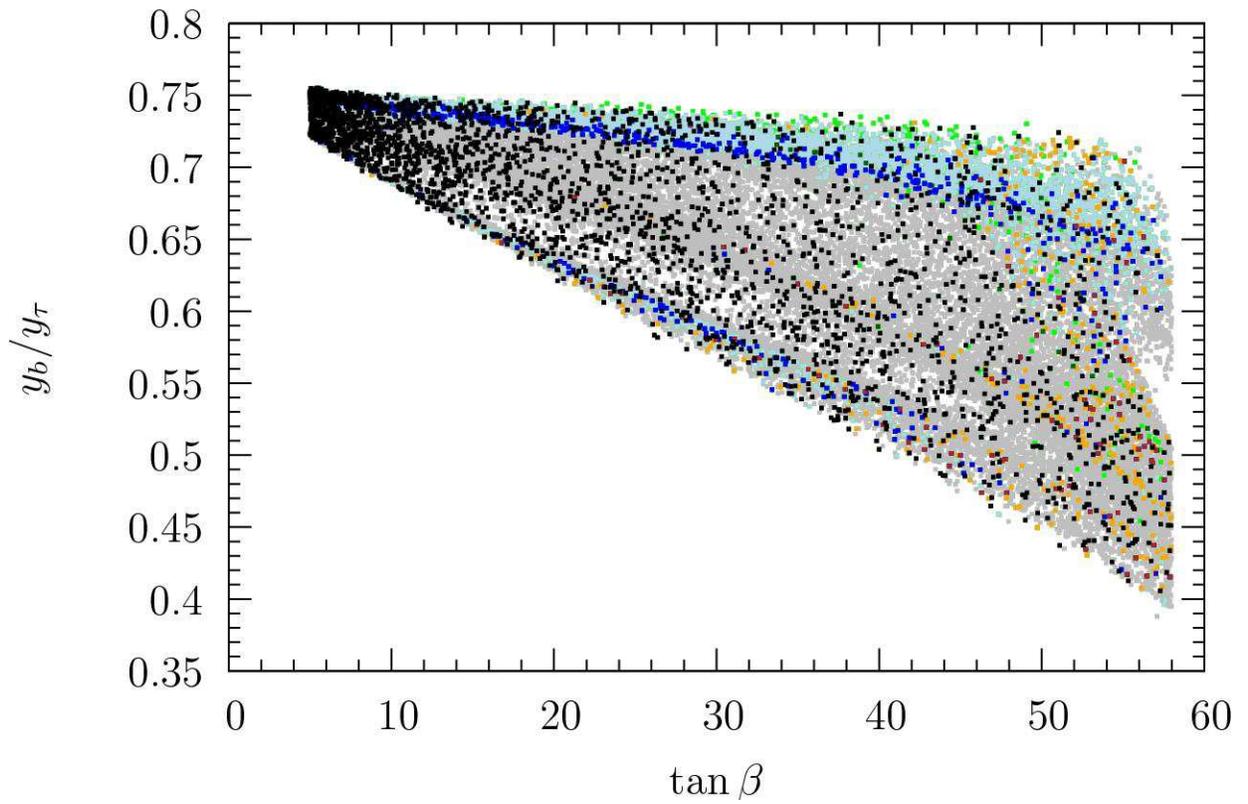}
\caption {Yukawa ratio $y_b/y_{\tau}$ (at $M_{\rm GUT}$) versus $\tan \beta$.
Color coding same as in Figure~\ref{fund-t30a0}.\label{yb_ybtau_tanb}}
\end{figure}

\begin{table}[t]
\begin{tabular}{c|cc|cc}
\hline
\hline
          & SU(5) & CMSSM & SU(5) & CMSSM    \\
\hline
$m_{1/2}$ &  780  & 780   &  788   &   788   \\
$m_{\bar 5} $    &  9.66 & 483   &  20.9  &  1047   \\
$m_{10} $ &  483  & 483   &  1047  &  1047   \\
$\tan\beta$ &  50  & 50   &  50   &   50\\
$A_0$ &  0  & 0   &  0   &   0   \\
\hline
$m_h$          & 117   & 117  & 118   & 117  \\
$m_H$          & 798   & 767  & 1032  & 879  \\
$m_A$          & 793   & 752  & 1026  & 874  \\
$m_{H^{\pm}}$  & 802   & 762  & 1036  & 884  \\
\hline
$m_{\tilde{\chi}^{\pm}_{1,2}}$
& 624, 990 & 624, 907 & 637, 1237 & 635, 885  \\
$m_{\tilde{\chi}^0_{1,2,3,4}}$ & 330, 623, 981, 989   & 330, 623,
896, 924
& 336, 636, 1232, 1236 & 336, 634, 873, 885   \\
$m_{\tilde{g}}$ & 1743 & 1748 & 1784 & 1796   \\
\hline $m_{{\tilde{u}}_{1,2}}$
& 1597, 1654 & 1597, 1655 & 1857, 1906 & 1857, 1905  \\
$m_{\tilde{t}_{1,2}}$
& 1286, 1506 & 1265, 1487 & 1483, 1721 & 1399, 1639  \\
\hline $m_{{\tilde{d}}_{1,2}}$
& 1511, 1656 & 1591, 1657 & 1512, 1907 & 1851, 1906  \\
$m_{\tilde{b}_{1,2}}$
& 1367, 1486 & 1412, 1482 & 1367, 1698 & 1593, 1662  \\
\hline
$m_{\tilde{\nu}_{1,2,3}}$
& 515, 515, 450 & 705, 705, 639 & 513, 513, 358 & 1166, 1166, 1030  \\
\hline
$m_{{\tilde{e}}_{1,2}}$
& 524, 563  & 563, 711 & 525, 1086 & 1086, 1169  \\
$m_{\tilde{\tau}_{1,2}}$
& 354, 511  & 338, 661 & 349, 957  & 750, 1030   \\
\hline
$\Omega_{CDM}h^2$ &  0.115 & 0.053  & 0.118   & 0.175  \\
\hline
$y_b/y_{\tau}$($M_{GUT}$) & 0.52  & 0.55  & 0.49   & 0.58  \\
\hline $\sigma_{\tilde{\chi}_1^0-p, {\rm SI}}({\rm pb})$ & $4.00
\times 10^{-10}$ & $6.00 \times 10^{-10}$ & $1.13 \times 10^{-10}$
& $4.64 \times 10^{-10}$ \\
$\sigma_{\tilde{\chi}_1^0-n, {\rm SI}}({\rm pb})$
& $4.29 \times 10^{-10}$
& $6.45 \times 10^{-10}$
& $1.21 \times 10^{-10}$
& $4.96 \times 10^{-10}$ \\
\hline
\hline
\end{tabular}
\caption{ Sparticle and Higgs masses (in units of GeV),
with $m_t=172.6$ GeV, $\tan \beta=50$, and $\mu>0$. We present two SU(5) benchmark points 
and the corresponding CMSSM points for comparison. Also included are the spin-independent 
neutralino-nucleon interaction cross-sections. Note that $2~y_b \approx y_{\tau}$ at $M_{\rm GUT}$.
} \label{table1}
\end{table}

\begin{table}[t]
\centering
\begin{tabular}{c|cc|c}
\hline
\hline
         & SU(5) & CMSSM & SU(5)     \\
\hline
$m_{1/2}$ &  287  &  287  &   275   \\
$m_{\bar 5} $    &  475  & 1203  &  73.9   \\
$m_{10} $ & 1203  & 1203  &   850   \\
$\tan \beta$ & 50 &    50 &    10   \\
$A_0$     & -1000 & -1000 & -2000   \\
\hline
$m_h$          & 115   & 115  &   119   \\
$m_H$          & 894   & 602  &  1130   \\
$m_A$          & 888   & 598  &  1123   \\
$m_{H^{\pm}}$  & 898   & 609  &  1133   \\
\hline
$m_{\tilde{\chi}^{\pm}_{1,2}}$
& 236, 1059    &  232, 525    & 222, 1124  \\
$m_{\tilde{\chi}^0_{1,2,3,4}}$
& 121, 236, 1055, 1058 & 120, 232, 515, 525 & 115, 223, 1110, 1121 \\
$m_{\tilde{g}}$ & 746  & 757  & 696 \\
\hline $m_{{\tilde{u}}_{1,2}}$
& 1336, 1342   & 1333, 1338  &  1026, 1036 \\
$m_{\tilde{t}_{1,2}}$
&  926, 1143   &  780, 984 &  366,  853 \\
\hline $m_{{\tilde{d}}_{1,2}}$
& 743, 1345   & 1334, 1341 &  571, 1039  \\
$m_{\tilde{b}_{1,2}}$
& 554, 1121   &  947, 1065 &  542, 811  \\
\hline
$m_{\tilde{\nu}_{1,2,3}}$
& 497, 497, 142& 1214, 1214, 1041 &   175, 175, 120 \\
\hline
$m_{{\tilde{e}}_{1,2}}$
& 506, 1208 & 1207, 1217 &   203, 857   \\
$m_{\tilde{\tau}_{1,2}}$
& 130, 998  & 816, 1045 &  151, 839   \\
\hline $\Omega_{CDM}h^2$
&  0.105 & 3.78  & 0.106  \\
\hline
$y_b/y_{\tau}$($M_{GUT}$) &  0.44  & 0.64  &  0.64     \\
\hline $\sigma_{\tilde{\chi}_1^0-p, {\rm SI}}({\rm pb})$ & $2.69
\times 10^{-10}$ & $2.80 \times 10^{-9}$
& $4.57 \times 10^{-11}$  \\
$\sigma_{\tilde{\chi}_1^0-n, {\rm SI}}({\rm pb})$
& $2.91 \times 10^{-10}$
& $3.04 \times 10^{-9}$
& $4.81 \times 10^{-11}$  \\
\hline
\hline
\end{tabular}
\caption{Sparticle and Higgs mass spectra (in units of GeV)
for two additional SU(5) benchmark points (compare Table~\ref{table1}), 
with $m_t=172.6$ GeV and $\mu>0$. The CMSSM equivalent of the first point is also included.
The CMSSM equivalent of the second point gives tachyonic solutions and is therefore 
omitted. Also included are the spin-independent
neutralino-nucleon interaction cross-sections. } \label{table2}
\end{table}

\end{document}